\documentclass[10pt,onecolumn]{article}
\usepackage{ifpdf}
\ifpdf
   \pdfoutput=1        %
   \pdftrue
\fi
\ifpdf
  \usepackage[pdftex]{geometry}
\else
  \usepackage{color}
  \usepackage{geometry}
\fi
\usepackage[english]{babel}
\usepackage[latin1]{inputenc}
\usepackage[T1]{fontenc}
\usepackage[normalem]{ulem}
\usepackage{pslatex}
\usepackage{setspace}
\usepackage{longtable}
\usepackage{multirow}
\usepackage[bookmarks=false]{hyperref}
\usepackage{authblk}

\hypersetup{
  pdfkeywords={}
}
\begin{document}
\title{Implementing Risk-Limiting Post-Election Audits in California}

\date{\textit{Note: the canonical version of this paper is available
    here:} \url{http://josephhall.org/papers/rla_evt09.pdf}}

\renewcommand\Affilfont{\small}

\author[1,2,*]{Joseph Lorenzo Hall}
\author[3]{Luke W. Miratrix}
\author[3]{Philip B. Stark}

\author[4]{Melvin Briones}
\author[4]{Elaine~Ginnold}
\author[5]{Freddie Oakley}
\author[6]{Martin Peaden}
\author[6]{Gail Pellerin}
\author[5]{Tom Stanionis}
\author[6]{Tricia~Webber}

\affil[1]{University of California, Berkeley; School of Information}
\affil[2]{Princeton University; Center for Information Technology Policy}
\affil[3]{University of California, Berkeley; Department of Statistics}
\affil[4]{Marin County, California; Registrar of Voters}
\affil[5]{Yolo County, California; County Clerk/Recorder}
\affil[6]{Santa Cruz County, California; County Clerk}

\maketitle

\begin{abstract}
  Risk-limiting post-election audits limit the chance of certifying an
  electoral outcome if the outcome is not what a full hand count would
  show.
  Building on previous work~\cite{stark_conservative_2008,
    stark_sharpdiscmeas_042008, stark_cast_022009,
    stark_pvalues_022009, miratrix_trinomial_032009}, we report
  pilot risk-limiting audits in four elections during 2008 in three
  California counties: one during the February 2008 Primary Election
  in Marin County and three during the November 2008 General Elections
  in Marin, Santa Cruz and Yolo Counties.
  We explain what makes an audit \textit{risk-limiting} and how
  existing and proposed laws fall short.
  We discuss the differences among our four pilot audits.
  We identify challenges to practical, efficient risk-limiting audits
  and conclude that current approaches are too complex to be used
  routinely on a large scale.
  One important logistical bottleneck is the difficulty of exporting
  data from commercial election management systems in a format
  amenable to audit calculations.
  Finally, we propose a bare-bones risk-limiting audit that is less
  efficient than these pilot audits, but avoids many practical
  problems.
\end{abstract}

\section{Introduction}
\label{sec:intro}

\let\oldthefootnote\thefootnote
\renewcommand{\thefootnote}{\fnsymbol{footnote}} 

\footnotetext[1]{To whom correspondence should be addressed. E-mail:
  \url{joehall@berkeley.edu}.  This paper will appear at the USENIX
  Electronic Voting Technology Workshop/Workshop on Trustworthy
  Elections (EVT/WOTE~'09) in Montreal, Canada, 10-11~August 2009.
  \textit{See:} \url{http://www.usenix.org/events/evtwote09/}. This is
  version 100 as of 10~July 2009.}

\let\thefootnote\oldthefootnote

Nearly a decade after the 2000 presidential election fiasco, the
``paper trail debate'' has all but ended: More and more jurisdictions
recognize that without indelible, independent ballot records that
reliably capture voter intent, auditing election outcomes is
impossible.
As auditable voting systems are adopted more widely, election
researchers are studying how to audit elections efficiently in a way
that ensures the accuracy of the electoral outcome.
The literature on the theory and practice of election auditing has
exploded recently: There have been nearly 70 papers and technical
reports since 2003.\footnote{Hall maintains an election audit
  bibliography~\cite{hall_auditbiblio}.}

Audits can be thought of as ``smart recounts'': Ideally, they ensure
accuracy the same way recounts do, but with less work.  Moreover,
audits can check the results of many contests at a time, not just
one contest on each ballot.
And audits can take place during the canvass period, before an
incorrect outcome is certified.
Audits help check the integrity of voting systems that use
computerized or electromechanical vote recording and tabulation
equipment.
The recent discovery that the election database of a voting system in
Humboldt County, California quietly dropped 197 ballots is a stark
reminder that examining audit records
is an important part of voting system
oversight~\cite{zetter_humboldt_122009}.

Election fraud using computerized voting systems appears to be rare,
and experts are hopeful that manual tally audits---as part of a
comprehensive election security plan---will detect and deter many
kinds of attacks~\cite{norden_bcscreview_082007}.
This would bolster and justify public confidence in the accuracy and
integrity of elections.

Indeed, several of the authors have been involved in improving
California's elections.
Hall served on the California Secretary of State's Top-To-Bottom
Review (TTBR)~\cite{casos_ttbr_2007} and has worked with hand tally
procedures~\cite{hall_audtprocs_072008}.
Ginnold and Stark served on the California Secretary of State's
Post-Election Audit Standards Working
Group~\cite{jefferson_peaswg_072007} (PEASWG).
Their experience made it clear that no existing audit method
controlled the risk of certifying an incorrect
outcome:\footnote{Throughout this paper, an ``incorrect,''
  ``erroneous'' or ``wrong'' apparent outcome is one that disagrees
  with the outcome that a full manual count of the audit trail would
  show.  If the audit trail is accurate and complete and the manual
  counting process is perfect, the outcome of a such a count shows how
  the votes were actually cast.  Obviously, there are many ways the
  audit trail could be less than perfect.
  Meticulous chain of custody is crucial.  And hand counting is
  subject to error.  Even so, the result of a hand count of the audit
  trail is generally the legal touchstone,
  the ``true'' outcome of the election.} There was no
method to decide whether it was safe to stop auditing---given the
discrepancies observed in the sample---or necessary to continue to a
full manual count.

Then-extant statistical methods for post election auditing focused on the following
question: If the apparent outcome of the election differs from the outcome a
full hand count would show, how big a sample is needed to ensure a high chance
of finding at least one error?
This ``detection'' paradigm makes sense in some contexts, for instance, if the voting technology is
direct-recording electronic machines (DREs) and the paper audit trail is perfect. 
Then, if even a single discrepancy between the DRE record and the paper were found, 
it would indicate a serious problem calling into question the outcome of the contest, 
and the entire paper audit trail should be examined. 

However, occasional discrepancies between a counting board's determination of voter intent 
and a machine reading of a voter-marked paper ballot are virtually inevitable.  
Audits of any modestly large number of voter-marked ballots will almost certainly 
find one or more discrepancies.  
What then?  
Since error was detected, should the entire audit trail be counted by hand? 

This suggests a different paradigm: risk-limiting audits.  
In the detection paradigm, we ask for a large chance of finding at least one error 
whenever the outcome is wrong. 
In the risk-limiting paradigm, we ask for a large chance of a full 
hand count whenever the outcome is wrong.
That shift is crucial.

To turn an audit procedure created in the detection paradigm into a risk-limiting audit 
requires a full manual count whenever the audit finds even a single error.  
It is preferable to start from scratch to develop risk-limiting methods,
methods that can stop short of a full hand count if the audit yields 
sufficiently strong evidence that the outcome is correct. 
(The strength of the evidence can be measured by a $P$-value; 
see~\cite{stark_pvalues_022009}.)
The detection question is, ``if the outcome is wrong, is there a big chance that the 
audit will find at least one error?''
The risk-limiting question is,``if the outcome is wrong, is there a big chance the audit would have 
found more error than it did find?''

Stark~\cite{stark_conservative_2008,stark_sharpdiscmeas_042008} was the first to develop 
risk-limiting audit methods.
Those methods work by collecting data, assessing
whether those data give strong evidence that the outcome is right, and
collecting more data if not.
The basic approach, with variations and refinements, was used in the
four audits reported here: the first ``live'' uses of risk-limiting
methods during a canvass to confirm electoral outcomes statistically,
before they are certified.

We hoped to answer several questions with these pilots:
What methods are practical for use during the post-election canvass
period?
What resources are required?
What challenges and opportunities do jurisdictions face if they
implement risk-limiting audits?

The paper is organized as follows: Section~\ref{sec:rla-bgrnd}
explains what risk-limiting audits are and what they are not, and
reviews current audit legislation in the United States.
Section~\ref{sec:rla-ca} describes the four pilot risk-limiting
audits.
Section~\ref{sec:discuss} discusses what these pilots revealed about
conducting risk-limiting audits.
Section~\ref{sec:modprop} proposes a very simple risk-limiting audit
that avoids some of the issues encountered in our pilot studies, but
is less efficient.
Section~\ref{sec:conc} concludes with some comments on future work.

\section{Risk Limiting Audits Defined}
\label{sec:rla-bgrnd}

This section explains what is and what is not a risk-limiting audit.
What distinguishes risk-limiting audits from other election audits is
that they have a big, pre-specified chance of catching \textit{and
  correcting} incorrect electoral outcomes.
The mechanism for correcting an incorrect outcome is a full hand
count; generally, it is not legal (nor a good idea) to alter the
apparent preliminary outcome on statistical grounds alone, because it
introduces the possibility that a correct apparent electoral outcome
would be rendered incorrect.
Instead, when there is not strong evidence that the apparent outcome
is right, a risk-limiting method progresses to a full hand count,
which---by definition---shows the right outcome.
Thus a risk-limiting audit either reports the apparent outcome, which
might be right or wrong, or the outcome of a full hand count, which
must be right.
The chance that a risk-limiting audit reports the outcome of a full
hand count is high if the apparent outcome is wrong.
When the apparent outcome is right, an efficient risk-limiting audit
tries to count as few ballots as possible to confirm the outcome.

\subsection{What they are}
\label{sec:what}

Risk-limiting audits are a special kind of post-election manual tally
(PEMT).
PEMTs check the accuracy of vote tabulation by comparing reported vote
subtotals for batches of ballots\footnote{A ``batch'' is an arbitrary
  grouping, but every ballot must be in exactly one batch.  For
  instance, a batch might consist of all ballots for a precinct cast
  in the polling place, and another batch might consist of all ballots
  for that same precinct cast by mail (absentee ballots).  Provisional
  ballots could comprise another batch.} with subtotals derived by
counting the votes in those batches by hand.
PEMTs are impossible unless:\footnote{Any voting system that captures
  an indelible, voter-verifiable audit record that can be sampled and
  counted independently could be audited using risk-limiting methods.
  The authors have limited experience with cryptographic
  and ``open-audit'' voting systems, but we believe risk-limiting
  audits of those systems are possible and desirable.
}
\begin{enumerate}
\item Vote subtotals are reported separately for the batches: There
  must be ``something to check.''  The subtotals must be reported
  before batches are selected for hand counting.
\item The ballots are available: There must be ``something to check
  against.''  They must be the same ballots that voters had the
  opportunity to verify and from which the tabulation process created
  the vote subtotals.
\item The batches of ballots are counted by hand: There must be ``an
  independent way to check'' the subtotals.
\end{enumerate}
Jurisdictions in 25~states are legally required to perform some type
of post-election manual tally.
We discuss differences among these PEMT schemes in
Section~\ref{sec:whatnot}.

Not every PEMT limits the risk of certifying an incorrect electoral
outcome.
Indeed, to the best of our knowledge, only four PEMTs have been
risk-limiting---the four audits we report here.
The consensus definition of a risk-limiting audit, endorsed by the
American Statistical Association and a broad spectrum of election
integrity advocates, is:
\begin{quote}
  Risk-limiting audits [are audits that] have a large, pre-determined
  minimum chance of leading to a full recount whenever a full recount
  would show a different outcome.~\cite{audit_principles_112008}
\end{quote}
The ``risk'' is the maximum chance that there is not a full count
if the outcome is incorrect.

There are many ways to implement risk-limiting audits.
By definition, all risk-limiting audits control the chance of stopping
short of a full hand count when the apparent outcome is wrong.
But they differ in their efficiency: the amount of counting they
require when the outcome is in fact correct.
Other types of audits---e.g., fixed-percentage audits, tiered audits
and polling audits,\footnote{Norden, Burstein, Hall and
  Chen~\cite{norden_bcscreview_082007} discuss these types of audits.}
do not keep the risk below any pre-determined level.
Indeed, such audits generally do not control risk at all.

A risk-limiting audit ends in one of two ways.
Either the audit stops before every ballot has been audited, or the
audit continues until every ballot has been counted by hand.
In the first case, a full hand count might have shown that the
apparent winner is not the true winner.
If so, an electoral error occurs.
In the second case, there is no chance of electoral error---the full
hand count shows the true winner, by definition.
The audit limits risk if it keeps the chance of making an electoral
error small when the apparent outcome is incorrect.
The audit is efficient if it does not count many ballots when the
apparent outcome is correct.
If the apparent outcome is wrong, the audit should count every
ballot---efficiency is not an issue.

So, to be a \textit{risk-limiting} audit, a PEMT must have an
additional element:
\begin{enumerate}
\setcounter{enumi}{3}
\item \label{item:ra-criterion} A minimum, pre-specified chance that,
  if the apparent outcome of the election is wrong,
  \textit{every} ballot will be tallied by hand.
\end{enumerate}
Any audit with element~\ref{item:ra-criterion} is risk-limiting, by
definition.
Risk-limiting audits generally have two more elements:
\begin{enumerate}
\setcounter{enumi}{4}
\item \label{item:ra-test} A way to assess the evidence that the
  apparent outcome is correct, given the errors found by the hand
  tally.
\item \label{item:ra-expand} Rules for enlarging the sample if the
  evidence that the apparent outcome is correct is not sufficiently
  strong.
\end{enumerate}
Elements~\ref{item:ra-test} and~\ref{item:ra-expand} allow the
procedure to work sequentially: Collect data, assess evidence, and
(i)~stop auditing if the evidence is strong that the outcome is right,
or (ii)~collect more data (expand the audit) if the evidence is not
sufficiently strong.
Testing sequentially can require far less counting when the apparent
outcome is correct.

In unpublished work, Johnson~\cite{johnson_vvpataudit_102004} appears
to be the first to have approached election auditing as a sequential
testing problem.
However, Johnson's approach relies on auditing individual ballots,
comparing electronic vote records directly with corresponding physical
audit records chosen at random.
Current voting systems do not support ``single-ballot audits,''
although there have been proposals for systems that would.

Stark and his collaborators have developed risk-limiting audits using
sequential tests based on comparing hand counts of randomly selected
batches of ballots with the reported results for the same
batches~\cite{stark_conservative_2008, stark_cast_022009,
stark_sharpdiscmeas_042008, stark_pvalues_022009,
miratrix_trinomial_032009, stark_collections_052009}.
Hand counts of randomly selected batches of ballots are the basis of
current and proposed auditing laws.

Stark's first treatment~\cite{stark_conservative_2008} addressed
simple random samples (SRS) and stratified random samples of batches,
which is how most jurisdictions with PEMTs select batches to audit.
He treated the data as a ``telescoping'' sample: At each stage, the
sample was considered to consist of all the data collected so far.
He found that a new measure of discrepancy between the machine and
hand count, the maximum relative overstatement of pairwise margins
(MRO), improved the efficiency
markedly~\cite{stark_sharpdiscmeas_042008}.
Instead of treating the sample as telescoping, one can condition on
errors found in previous audit stages~\cite{stark_cast_022009}.
This allows a rigorous treatment of ``targeted''
auditing---deliberately sampling some batches of ballots---which also
can improve efficiency.

Stark~\cite{stark_pvalues_022009} and Miratrix and
Stark~\cite{miratrix_trinomial_032009} developed risk-limiting audits
using more efficient sampling designs: sampling with probability
proportional to error bounds (PPEB) and the negative exponential
(NEGEXP) sampling method of Aslam, Popa and
Rivest~\cite{aslam_auditing_072008}.
Financial and electoral audits have much in common, including the fact
that errors are typically zero or small, but can be large---which can
make parametric approximations very inaccurate.
PPEB sampling is common in financial auditing, where the error bound
is the reported dollar value of an account.
The trinomial bound method of Miratrix and
Stark~\cite{miratrix_trinomial_032009} is closely related to the
multinomial bound method, one of several used in financial auditing to
analyze PPEB samples.

Stark~\cite{stark_collections_052009} extended MRO to get a combined
measure of error for a collection of races.
That makes it possible to perform a risk-limiting audit of several
races simultaneously, with less effort than would be required to audit
them separately.
In work in progress, Miratrix and Stark use the Kaplan-Markov
Martingale approach described by Stark~\cite{stark_pvalues_022009} to
implement much more efficient sequential tests.

\subsection{What they are not}
\label{sec:whatnot}

This section discusses audit legislation and a pilot audit in Boulder
County, CO.
As far as we are aware, no proposed or enacted legislation mandates a
risk-limiting audit, according to the consensus definition given in
section~\ref{sec:what}, and no audits other than the four reported
below in section~\ref{sec:rla-ca} have been risk-limiting.

Audits and PEMT laws generally have focused on how large an audit
sample to start with.
That is important, but not as important as having a sound way to
decide whether to stop counting or to enlarge the sample after the
initial sample has been audited.
If an audit procedure does not guarantee a known
minimum probability
of a full hand count whenever the electoral outcome is
wrong, the audit is not risk-limiting.
The initial sample size is not important for controlling the
risk\footnote{The initial sample size can affect the efficiency,
  though.} as long as there is a proper calculation of the strength of
the evidence that the outcome is correct, and the audit is expanded if
the evidence is not strong---eventually to a full manual count.

Heuristically, the evidence that the outcome is correct is weak if
the sample size is small, if the margin is small, or if the initial
audit finds too many errors.
The difficulty is in making these heuristics precise---the problem
addressed by the various papers on risk-limiting
audits~\cite{stark_conservative_2008, stark_cast_022009,
stark_sharpdiscmeas_042008, stark_pvalues_022009,
miratrix_trinomial_032009, stark_collections_052009}.
As illustrated in section~\ref{sec:rla-ca}, efficient risk-limiting
methods have unavoidable complexity that might make them unsuitable
for broad use, although we are hopeful that better ``data plumbing''
will help.

\subsubsection{Existing State Legislation}
\label{sec:stateleg}

The most common prescription for PEMT audits involves selecting a
pre-determined percentage of batches of ballots (e.g., precincts,
machines, districts), counting the votes in those batches, and
stopping.\footnote{\label{fn:2}The authors are aware of the following
  state-level post-election audit provisions that use tiered- or
  fixed-percentage audit designs: Alaska specifies one precinct per
  election district that must consist of at least 5\% of ballots cast
  (Alaska Stat.\ \S~15.15.430 (2009)); Arizona specifies the greater
  of two percent of precincts or two precincts (A.R.S.\ \S~16-602
  (2008)); California specifies 1\% of precincts (Cal Elec Code
  \S~15360 (2008)); Colorado specifies no less than 5\% of voting
  devices (C.R.S.\ 1-7-514 (2008)); Connecticut specifies no less than
  10\% of voting districts (Conn.~Gen.~Stat.\ \S~9-320f (2008));
  Florida specifies no less than 1\% but no more than 2\% for one
  randomly-selected contest (Fla.~Stat.\ \S~101.591 (2009)); Hawaii
  specifies no less than 10\% of precincts (HRS \S~16-42 (2008));
  Illinois specifies 5\% of precincts (10~ILCS~5/24A-15 (2009))
  (allows machine retabulation); Kentucky specifies between 3--5\% of
  the number of total ballots cast (KRS \S~117.383 (2009)); Minnesota
  specifies 2 precincts, 3 precincts, 4 precincts or at least 3\% of
  precincts per jurisdiction, depending on the number of registered
  voters (Minn.~Stat.\ \S~206.89 \textit{et seq.} (2008)); Missouri
  specifies in its state administrative rules the greater of 5\% of
  precincts or one precinct (15~CSR~30-10.110(2)); Montana specifies
  at least 5\% of precincts and at least one federal office,
  statewide office, statewide legislative office, and one statewide
  referendum (2009~Mt.  SB~319); Nevada specifies in administrative
  rules between 2--3\% depending on the jurisdiction's population
  (Nevada Administrative Code, Ch.~293.255) (allows machine
  retabulation); New Mexico specifies 2\% of voting systems
  (N.M.~Stat.~Ann.\ \S~1-14-13.1 (2008)) (\textit{see} further
  discussion in:~\ref{sec:stateleg}); New York specifies 3\% of voting
  machines (NY~CLS~Elec \S~9-211 (2009)); Oregon specifies a tiered
  audit structure of 3\%, 5\% or 10\% of precincts depending on the
  margin of the contest (ORS \S~254.529 (2007)); Pennsylvania
  specifies the lesser of 2000 or 2\% of votes (25~P.S.~\S~3031.17
  (2008)) (allows machine retabulation); Tennessee specifies at least
  3\% of votes and at least 3\% of precincts (Tenn.~Code~Ann.\ 
  \S~2-20-103 (2009)); Texas specifies the greater of 3 precincts or
  1\% of precincts (Tex.~Elec.~Code \S~127.201 (2009)); Utah specifies
  at least 1\% of machines (\textit{see:}
  \S~6~of~\cite{utah_electionpolicy_2006}); Washington specifies up to
  4\% machines (Rev.~Code~Wash.\ (ARCW) \S~29A.60.185 (2009)) (only
  1\% is required to be counted by hand); West Virginia specifies 5\%
  of precincts (W.~Va.~Code \S~3-4A-28 (2008)); Wisconsin specifies
  5~``reporting units'' for each voting system
  (\textit{see:}~\cite{WI_auditreqs_2006} implementing Wis.~Stat.\ 
  \S~7.08(6) (2008)) (audit occurs only after each General Election).
  The following states' audit laws do not require auditing of all
  contests on the ballot: Arizona, Connecticut, Florida, Minnesota,
  Missouri, Montana, Tennessee, Washington and Wisconsin.  The
  District of Columbia recently issued an emergency rule requiring
  manual audits of 5\% of precincts (\textit{see:}
  \cite{vvf_manualauditreqs_2009} at~4).  Vermont has no legal
  requirement for manual audits but the Secretary of State may order
  them under certain conditions (17~V.S.A. \S~2493 (2009)).  Ohio
  Secretary of State ordered a 5\% manual audit for the November 2008
  General Election using her power of Directive (\textit{See:}~\cite{oh_sos_dir-2008-113}).
  The Verified Voting Foundation (VVF) maintains a useful and
  regularly-updated dossier of these
  provisions~\cite{vvf_manualauditreqs_2009}.}
A notable exception is North Carolina, where the manual audit statute
requires the audit sample size to be ``chosen to produce a
statistically significant result and shall be chosen after
consultation with a statistician.''\footnote{N.C.~Gen.~Stat.\ 
  \S~163-182.1--182.2 (2009).}
Unfortunately, this is a misuse of the term of art ``statistically
significant.''
The wording does not make sense to a statistician.

New Jersey's PEMT audit law\footnote{N.J.~Stat. \S~19:61-9 (2009).}
tries to enunciate risk-limiting audit principles; indeed, a co-author
of this legislation claims it is ``risk-based.''\footnote{Stanislevic
calls the N.J. law the first ``risk-based statistical audit law.''
\textit{See:} Howard Stanislevic, ``Election Integrity: Fact \&
Friction'', at: \url{http://e-voter.blogspot.com/}.}
The statute creates an ``audit team'' to oversee manual audits of
voter-verified paper records and requires that the procedures the team
adopts:
\begin{quote}
  \dots ensure with at least 99\% statistical power that for each
  federal, gubernatorial or other Statewide election held in the
  State, a 100\% manual recount of the voter-verifiable paper records
  would not alter the electoral outcome reported by the audit\dots
  \footnote{N.J.~Stat.\ \S~19:61-9(c)(1) (2009).}
\end{quote}
This misuses the statistical term of art ``power'': The language
does not make sense to a statistician.
Since New Jersey's current voting equipment does not produce an audit
trail, the New Jersey audit law cannot help ensure
accuracy.\footnote{As in New Jersey, manual audits are required
  by law in Kentucky and Pennsylvania but neither state requires
  auditable voting systems.  Depending on the type of voting
  technology, there may or may not be anything to count by hand.}

The New Jersey statute goes on to say that auditors may adopt
``scientifically reasonable assumptions,'' including:
\begin{quote}
  \dots the possibility that within any election district up to 20\%
  of the total votes cast may have been counted for a candidate or
  ballot position other than the one intended by the voters.\dots
  \footnote{\textit{Id.}}
\end{quote}
This assumption is sometimes called a within-precinct-miscount or
within-precinct-maximum (WPM) bound.\footnote{The term ``WPM''
  suggests that the audit unit is a precinct, but often the term is
  used more broadly to denote an upper bound on the number of errors
  in an auditable batch as a percentage of the reported number of
  ballots or votes in the batch.  ``WBM'' (within-batch-miscount)
  might be a better term.}
The New Jersey rule corresponds to a WPM of 20\%.

The chance that a random sample will find one or more batches with
error depends on the number of batches that have error: the more
batches with errors, the greater the chance.
The number of batches that must have errors for the apparent electoral
outcome to be wrong depends on the amount of error each batch can hold
(and on the margin).
If batches can hold large errors, few batches need to have errors for
the outcome to be wrong.

WPM limits the amount of error that each batch can hold---by
assumption.
WPM implies that if there is enough error to change the
outcome, the error cannot be ``concentrated'' in very few batches:
There is a minimum number of batches that must have error if the
apparent outcome is wrong.
In turn, that implies that if the outcome is wrong, a sample of a
given size has a calculable minimum chance of finding at least one
batch with an error.
If the WPM assumption fails, however, outcome-changing error can hide in fewer
batches.
Then the chance that a sample of a given size finds a batch with
errors is smaller than the WPM calculation suggests: The chance of
noticing that there is something wrong is smaller than claimed.

We find WPM assumptions neither reasonable nor defensible.
There is no empirical or theoretical support for the assumption that
no more than 20\% of ballots in a batch can be counted incorrectly,
nor that an error of more than 20\% would always be caught without an
audit.
In fact, there is evidence to the contrary, including the recent
experience in Humboldt County, mentioned above, where 100\% of the
ballots in a batch were omitted.\footnote{The Humboldt case was not
  detected by a PEMT audit.  However, it proves
  that error can affect every ballot in a batch and yet go undetected
  during the canvass.}
The WPM assumption generally understates the amount of error that an
auditable unit can contain.\footnote{A 20\% bound on error can be
  optimistic or conservative, depending on whether there has been an
  accounting of ballots and depending on the distribution of reported
  votes---even within a single jurisdiction.  Typically, however, it
  is optimistic.}
Because WPM is not rigorous and tends to be optimistic, audits that
rely on WPM tend to understate the true risk, creating a false sense
of security.

Three other recently proposed laws are similar to the New Jersey
legislation.
New Mexico State Senate Bill~72, recently signed into law, has
language that sounds risk-limiting: It requires the sample to ensure
with ``at least ninety percent probability [\dots ] that faulty
tabulators would be detected if they would change the outcome of the
election for a selected office.''
Faulty tabulators are not the only reason apparent outcomes can be
wrong.
And the word ``detected'' is a problem.\footnote{It is not the only
  problem with the New Mexico law: The law ``hardwires'' sample sizes
  in a look-up table that appears to depend on a WPM-like error bound
  based on a snapshot of New Mexico precinct sizes.  The final text of
  SB~72 is available here:
  \url{http://www.nmlegis.gov/Sessions/09\%20Regular/final/SB0072.pdf}.
  This bill was signed into law by New Mexico Governor Richardson on 7
  April 2009.  \textit{See:}
  \url{http://www.governor.state.nm.us/press/2009/april/041009_07.pdf}.
  The law has not, at the time of writing, been codified into New
  Mexico's Election statutes (N.M.~Stat.~Ann. \S~1-13 \textit{et
    seq.}).}
There is a big difference between detecting error and determining that
the aggregate error might be large enough to change the apparent
electoral outcome; detecting error and requiring a full hand count are
not the same.
An audit does not limit risk unless it leads to full hand count
whenever there is less than compelling evidence that the apparent
outcome is correct---regardless of the reason the evidence is not
strong.
Most laws have no provision for expanding the audit even if the audit
uncovers large errors.

Massachusetts Senate Bill~356, and its companion House Bill~652, have
what appears to be good risk-limiting language.\footnote{\textit{See:}
  Massachusetts S.B.~356:
  \url{http://www.mass.gov/legis/bills/senate/186/st00pdf/st00356.pdf};
  Massachusetts H.B.~652:
  \url{http://www.mass.gov/legis/bills/house/186/ht00pdf/ht00652.pdf}.}
The Senate Bill states: ``\dots the audit shall be designed and
implemented to provide approximately a 99\% chance that a hand recount
of 100\% of the ballots will occur whenever such a recount would
reverse the preliminary outcome reported by the voting
system.''\footnote{\textit{Id.}  This is the risk-limiting language
  specific to statewide contests; for congressional races the
  probability is lowered to 90\%.}
The term ``approximately'' is not defined; it is unclear how much
deviation from the target probability is tolerable.
The bill has other problems, too: It does not audit all races and it
relies on a 25\% WPM assumption.
The House bill is much better: It does not use the ``approximately''
language, nor does it involve any WPM assumption.

Maryland House of Delegates Bill HB~665 appears similar to the New
Mexico bill.\footnote{It also tabulates sample sizes, but the table is
  more detailed.}
It lacks language comparable to the risk language in the New Jersey
and New Mexico laws.\footnote{This bill appears to have received no
  further action after its first reading.  \textit{See:}
  \url{http://mlis.state.md.us/2009rs/billfile/HB0665.htm}.}

\subsubsection{Emerging State Legislation}
\label{sec:emerg-state-legis}
 
Some state legislation and regulation come closer to mandating
features of risk-limiting audits.
Alaska, California, Hawaii, Minnesota, New York, Oregon, Tennessee,
and West Virginia hand count additional precincts or machines, in some
cases potentially to a full count, depending on the error found during
the audit.
Colorado recently passed an audit law that almost
requires a risk-limiting audit.
In this section we discuss the differences among these state-level
schemes.

Five of these States---Alaska, Hawaii, Oregon, Tennessee, and West
Virginia---have audit laws that can escalate to a full
count, but they do so using fairly blunt methods:

\begin{itemize}
\item Alaska requires counting one randomly selected precinct from
  each election district within the state.\footnote{\textit{Id.},
    note~\ref{fn:2}.}
  If the audit finds discrepancy amounting to 1\% between the hand
  count and the preliminary results, the audit expands to all ballots.
\item Hawaii requires an audit of 10\% of
  precincts.\footnote{\textit{Id.}, note~\ref{fn:2}.}
  If the audit finds any discrepancy, the law requires election
  officials to conduct an ``expanded audit''; however, the extent of
  the expanded audit is not specified.
\item Oregon requires a tiered initial audit of the ballots in 3\%,
  5\% or 10\% of precincts where the margin in a given race is greater
  than 2\%, between 1\% and 2\% or less than 1\%,
  respectively.\footnote{\textit{Id.}, note~\ref{fn:2}.}
  If the audit finds discrepancy between the hand count and the
  preliminary results of 0.5\% or more, the count has to be conducted
  again.
  If this level of discrepancy is confirmed by the second count, all
  ballots counted by the voting system on which these ballots were
  cast within the jurisdiction are counted.
\item Tennessee requires a hand count of 3\% of
  precincts.\footnote{\textit{Id.}, note~\ref{fn:2}.}
  If the difference between the hand count and electronic results is
  more than 1\%, the audit is expanded by an additional 3\% of
  precincts.
  Unfortunately, if the expanded audit still finds error amounting to
  a 1\% difference, the law here only ``authorizes'' the election
  officials to count additional precincts as they ``consider
  appropriate.''
\item West Virginia requires a manual count of VVPAT records in 5\% of
  precincts.\footnote{\textit{Id.}, note~\ref{fn:2}.}
  When the resulting hand count differs from the electronic results by
  more than one percent or when it results in a different outcome, the
  law requires all VVPAT records to be manually counted.
\end{itemize}

California, where we performed the audits described in this paper and
in other work~\cite{miratrix_trinomial_032009, stark_pvalues_022009,
hall_audtprocs_072008}, has regulations that expand the hand count if
enough error is found during the audit.
For almost 45 years, California has had a PEMT that audits a random
sample of 1\% of precincts.\footnote{ \textit{Id.}, note~\ref{fn:2}.
  In small races, the law can require auditing substantially more than
  1\% of precincts because it calls for auditing at least one precinct
  in every race.  For instance, a 4~precinct race would have at least
  1~precinct audited, resulting in at least a 25\% audit.  The new
  California PEMT regulations~\cite{pemt_ccrov09061_2009}, discussed
  in the text, call for a 100\% manual tally of all ballots cast on
  DRE voting systems.}
In the wake of studies by the Secretary of State's Top-To-Bottom
Review~\cite{casos_ttbr_2007} and Post-Election Audit Standards
Working Group~\cite{jefferson_peaswg_072007}, additional auditing
requirements were imposed in 2007 as a condition of recertification
for electronic voting systems.
The new rules were challenged in court and the Secretary has since
issued the Post-Election Manual Tally
Regulations~\cite{pemt_ccrov09061_2009} as emergency regulations.
Although the emergency rules are not risk-limiting, they have the
right flavor: They require more auditing for close contests and they
expand the audit---potentially to a full hand count---if the audit
uncovers many errors that overstated the margin.

Jurisdictions in Minnesota must tally votes in 2, 3 or 4~precincts, or
3\% of precincts, depending on the number of registered voters in the
jurisdiction.\footnote{\textit{Id.}, note~\ref{fn:2}.  Jurisdictions
  with more than ``100,000 registered voters must conduct a review of
  a total of at least four precincts, or three percent of the total
  number of precincts in the county, whichever is greater.''
  (Minn.~Stat.  \S~206.89(2)).}
Minnesota law says the audit must escalate by three precincts if it
``reveals a difference greater than one-half of one percent, or
greater than two votes in a precinct where 400 or fewer voters cast
ballots.''\footnote{Minn.~Stat.~206.89(a) (2008).}
If this first escalation finds a similar or greater amount of error
in the same jurisdiction,
the audit then escalates to encompass all precincts in the county.
As a third and final escalation step, the Secretary of State must
order a full recount of any race where results appear to be incorrect,
after these two stages of escalation, if these errors occurred in
counties that compromise more than ten percent of the vote count, in
aggregate.\footnote{Minn.~Stat.~206.89(b) (2008).}
These elements of the Minnesota law reduce risk: If enough error is
found during the hand count, the audit can grow to encompass the
entire race, even in races that cross jurisdictional boundaries.
However, the resulting risk still can be quite high, because
the law does not take sampling variability into account, because it
requires finding large errors in several precincts in each jurisdiction, and 
because the sampling fractions and escalation thresholds are fixed,
even for contests with very small margins.

New York's audit laws require the New York State Board of Elections to
promulgate regulations that determine when to increase the number of
voting systems in the audit and when to do a full count of the audit
records for all voting systems.\footnote{\textit{Id.},
  note~\ref{fn:2}.}
These regulations are currently available for public comment and
review.\footnote{\textit{See:} ``Proposed Amendment to Subtitle V of
  Title 9 of the Official Compilation of Codes, Rules and Regulations
  of the State of New York Repealing Part 6210.18 and Adding thereto a
  new Part, to be Part 6210.18 Three-Percent (3\%) Audit'', New York
  State Board of Elections, 29~May 2009,
  \url{http://www.elections.state.ny.us/NYSBOE/Law/6210.18Regulations.pdf}.}
The proposed regulations require a 3\% audit of all voting systems and
trigger an expanded audit of the records from an additional 5\% if any
vote share changes by 0.1\% or if an error occurs in at least 10\% of
machines in the initial sample.
The audit then expands in a similar manner to include paper records
from and additional 12\% and then finally encompasses all machines.

Each of these states has provisions for enlarging audits to a full hand
tally, depending on the frequency and location of errors the audit finds.
California, New York, and Minnesota tend to reduce risk---although not
to any pre-specified level and not for every contest.\footnote{While
  these provisions tend to reduce risk, they are not
  risk-limiting: California's regulation only triggers increased
  auditing when the margin of victory is less than 0.5\%.
  Contests with larger margins of victory are not subject to auditing
  beyond the standard 1\% PEMT audit, no matter how much error the 1\%
  audit finds.  Minnesota's law only audits races for U.S.\ President
  (or the Minnesota Governor), U.S.\ Senator and U.S.\ Representative.
  No other contests on the ballot are subject to the audit.  New
  York's proposed regulation does not coordinate audits across
  jurisdictional boundaries for contests that span multiple counties
  to limit the risk of certifying an incorrect outcome.  
  New York does not require
  escalation to a full count across all types of voting technology
  used to cast ballots in a contest, but instead confines escalation
  to the specific voting technology in which errors are observed.}

Finally, Colorado recently passed legislation that comes close to
mandating risk-limiting audits.
HB~1335 requires all counties to conduct what it calls
``risk-limiting'' audits by 2014, and establishes a pilot program to
develop procedures and regulations.\footnote{HB 09-1335, ``Concerning
  Requirements for Voting Equipment'', \textit{See:}
  \url{http://www.leg.state.co.us/Clics/CLICS2009A/csl.nsf/fsbillcont3/25074590521F41DA87257575005F1422?Open&file=1335_enr.pdf}.
  HB~1335 was signed into law by Colorado Governor Ritter on 15~May
  2009 (\textit{see:}~\cite{co_hb1335_ritterPR}).\label{fn:3}}
HB~1335 defines ``risk-limiting audit'' as:
\begin{quote}
  ``risk-limiting audit'' means an audit protocol that makes use of
  statistical methods and is designed to limit to acceptable levels
  the risk of certifying a preliminary election outcome that
  constitutes an incorrect outcome.\footnote{\textit{Id.},
  note~\ref{fn:3}.}
\end{quote}
This language comes closer to limiting the risk of certifying an
incorrect outcome than do the proposals discussed in the previous
section.

However, it has problems.
The phrase ``statistical methods'' serves to obfuscate, not clarify;
``risk'' is not defined, and the definition of ``incorrect outcome''
given in the statute has a loophole:
\begin{quote}
  ``incorrect outcome'' means an outcome that is inconsistent with the
  election outcome that would be obtained by conducting a full
  recount.\footnote{\textit{Id.}, note~\ref{fn:3}.}
\end{quote}
``Full recount'' might allow machine re-tabulation in lieu of a
full hand count of voter-verified ballot records---a more appropriate
standard for determining the ``correct'' electoral outcome.
Hence, a better legislative definition of ``risk-limiting audit'' is:
\begin{quote}
  ``risk-limiting audit'' means an audit protocol that has an
  acceptably high probability of requiring a full manual count
  whenever the electoral outcome of a full manual count would differ
  from the preliminary election outcome.  When the audit results in a
  full manual count, the outcome of that count shall be reported as
  the official outcome of the contest.
\end{quote}
That would be consistent with the consensus definition of
``risk-limiting audit,'' and still leave room for legislators or
elections officials to decide what ``acceptably high'' means.

\subsubsection{Federal Legislation}
\label{sec:federal-legislation}

Representative Rush Holt's ``Voter Confidence and Increased
Accessibility Act'' (H.R.~2894) is the leading federal election reform
bill to include PEMT audits.\footnote{\label{fn:4}H.R.~2894, ``The
  Voter Confidence and Increased Accessibility Act'', 111th
  U.S.~Congress (2009),
  \url{http://thomas.loc.gov/cgi-bin/bdquery/z?d111:h2894:} (accessed
  Jun 18, 2009).}

Like Oregon's legislation,\footnote{\textit{Id.},
  note~\ref{fn:2}.} the Holt bill has a tiered, margin-dependent
sample size of 3\%, 5\% or 10\% of precincts when the margin in
federal races is greater than 2\%, between 1\% and 2\% or smaller than
1\%, respectively.
The bill allows escalation---but does not require it---if errors are discovered during the audit.
Because the audit need not progress to a full hand count even when
large errors are found, the Holt bill does not limit risk.

The Holt bill has a clause that allows the National Institute of
Standards and Technology (NIST) to approve an alternative audit plan,
provided NIST determines that:
\begin{quote}
  (A) the alternative mechanism will be at least as statistically
  effective in ensuring the accuracy of the election results as the
  procedures under this subtitle; or
  
  (B) the reported election outcome will have at least a 95 percent
  chance of being consistent with the election outcome that would be
  obtained by a full recount.\footnote{\textit{Id.}, note~\ref{fn:4}.}
\end{quote}
This language has problems.
The Holt bill never requires a full hand count, so it cannot
ensure the accuracy of election results.
In particular, there is no sense in which it is ``statistically
effective in ensuring the accuracy of election results.''
It would seem that to approve an alternative under (A), NIST must
concede that the Holt bill is not statistically effective.

Clause (B)~looks more like a risk-limiting audit provision, but it is
garbled to a statistician's eye.
Absent another definition, we assume that ``reported election
outcome'' means ``apparent election outcome.''
The apparent outcome either is or is not the outcome a full recount
would show.
There is no probability about it.
The probability is only in the audit sample.
So, clause (B)~does not make sense.

Moreover, requiring ``consistency'' between the apparent outcome and
what a full recount would show seems too weak: It appears to permit an
apparent outcome to be altered without a full hand count.
If so, there is a possibility that a correct outcome will be turned
into an incorrect outcome based on statistical evidence.
That seems like it should be unacceptable.
These problems could be avoided by using the consensus definition of a
risk-limiting audit: The alternative mechanism should have at least a
95\% chance of requiring a full hand count whenever that hand count
would show that the apparent outcome was wrong.

We hope that if the Holt bill passes, the NIST clause will be
interpreted to allow risk-limiting audits.
Unfortunately, it is not clear that audits that satisfy the Holt
provisions can be risk-limiting.

\subsubsection{Boulder County, CO Audit, November 2008}
\label{sec:boulderaudit}

For the November 2008 General Election in Boulder County, Colorado,
the Boulder County Elections Division was assisted by McBurnett in
performing what he called a ``risk-limiting''
audit~\cite{mcburnett_boulder_112008}.
However, it is not risk-limiting according to the consensus 
definition.\footnote{\textit{See:} Section~\ref{sec:what}.}
It was designed in the ``detection'' paradigm, 
not the ``risk-limiting'' paradigm.

Under the assumption that WPM of 20\% holds (an assumption we find unconvincing), the 
Boulder County audit had a large chance of finding one or more errors if the outcome
were wrong---in local races, since errors in other counties were invisible to the audit.  
The number of batches to be audited for local races was capped at~10, so
the chance of finding at least one error if the outcome was wrong differed from 
local contest to local contest, depending on the margin, among other things.
The 10-batch limit was imposed so that auditing a
close, small contest would not require hand counting the votes of
every batch of ballots in the race.\footnote{In personal communication,
  McBurnett describes this as having had a ``fixed audit budget'' and
  that they chose to allocate that budget more towards larger
  contests.}

The Boulder audit did not have escalation rules---provisions for what to do 
if error was found.
Hence, it did not ensure any chance of a full hand count
if the apparent outcome was wrong.
The audit was constructed so if the outcome were wrong, there was a
large chance of finding at least one error.
The audit did find error in some contests.
Given the design, to be risk-limiting the audit had to escalate to a complete
hand count of every race in
which the initial sample found one or more errors, even assuming WPM of 20\% held.

\section{Risk-Limiting Audits in California}
\label{sec:rla-ca}

We performed four risk-limiting audits in California in 2008: two in
Marin County and one each in Yolo and Santa Cruz Counties.
This section describes the audits and the differences among
them.
Table~\ref{tab:audit-sum} reports summary statistics for the audits.
These audits are, to the best of our knowledge, the first and only
risk-limiting post-election audits, according to the consensus
definition discussed in Section~\ref{sec:what}.

The four audits explored different sampling methods, different
statistical tests, and a variety of administrative protocols to
increase efficiency.
They had a 75\% chance of leading to a full hand count, thereby
correcting an erroneous apparent outcome, if the apparent electoral
outcome happened to be wrong---no matter what caused the errors that
led to the incorrect outcome.
That is, these audits limited the risk that an incorrect outcome would
go uncorrected to at most 25\%.
We could have limited the risk to a lower level, at the cost of more
hand counting.
Because the primary goal of these audits was to gain experience,
compare methods, and to understand (and reduce) the logistical
complexity of administering risk-limiting audits, we felt that a risk
limit of 25\% was appropriate.

\begin{table}[t]
\begin{center}
{\small
\begin{tabular}{l|rrrrrrrrr}
County & Total   & Winner & Loser & Margin & Precincts & Batches  & Batches & \# Ballots  & \% Ballots\\
       & Ballots &        &       &        &           &          & Audited & Audited     & Audited \\
        \hline
Marin (A)  & 6,157   &  4,216 &  1,661 &  5.1\%  & 9   & 18  & 12 & 4,336 & 74\% \\
Yolo (W)   & 36,418  & 25,297 &  8,118 & 51.4\%  & 57  & 114 & 6  & 2,585 & 7\% \\
Marin (B)  & 121,295 & 61,839 & 42,047 & 19.1\%  & 189 & 544 & 14 & 3,347 & 3\% \\
Santa Cruz & 26,655  & 12,103 &  9,946 &  9.6\%  & 76  & 152 & 16 & 7,105 & 27\%
\end{tabular}
}
\end{center}
\caption{\protect
  Summary of the four races audited.
  Ballots and votes for the candidates are the results as reported when the audit 
  commenced; they may differ from the official final results for the contests.  
  Margins are expressed as a percentage of the votes for all candidates.
  Marin Measure~A required a {\small
  $2/3$}~supermajority to pass (the margin is calculated accordingly from 5,877 total valid ballots).
  Yolo County Measure~W required a simple majority.  
  Marin Measure~B required a simple majority.  
  These three measures passed.
  John Leopold and Betty Danner were the main contenders for 
  Santa Cruz County Supervisor, 1st~District; 
  there were 103~votes in all for write-in candidates.
  Leopold won.
}
\label{tab:audit-sum}
\end{table}

\subsection{Marin County, Measure A, February 2008}
\label{sec:marin-feb08}

\begin{table}[t]
\begin{center}  
\begin{tabular}{lrr|rrr}
Batch ID & $b_p$ & Bound & Yes & No & Audited \\
\hline
2001-IP    & 391 & 286 & 278 & 101 & yes \\
2001-VBM   & 657 & 456 & 438 & 193 & no\\
2004-IP    & 284 & 214 & 204 & 66  & yes\\
2004-VBM   & 389 & 268 & 257 & 116 & yes \\
2010-VBM   & 6   & 4   & 4   & 2   & no\\
2012-IP    & 218 & 173 & 167 & 43  & yes\\
2012-VBM   & 342 & 250 & 242 & 89  & no\\
2014-IP    & 299 & 221 & 214 & 75  & no\\
2014-VBM   & 420 & 319 & 306 & 95  & yes\\
2015-IP    & 217 & 171 & 167 & 44  & yes\\
2015-VBM   & 483 & 346 & 332 & 131 & yes\\
2019-IP    & 295 & 222 & 215 & 70  & yes\\
2019-VBM   & 567 & 403 & 395 & 160 & yes\\
2101-IP    & 265 & 181 & 169 & 79  & no\\
2101-VBM   & 439 & 296 & 275 & 133 & yes \\
2102-IP    & 223 & 152 & 144 & 68  & yes\\
2102-VBM   & 410 & 257 & 233 & 142 & yes\\
ALL-PRO    & 252 & 191 & 176 & 54  & no\\
\end{tabular}
\end{center}
\caption{\protect 
Results and error bounds for Marin Measure~A, February~2008.  
A stratified random sample of 12~batches was selected by rolling
10-sided dice.  Batch ID is the precinct number followed by the
manner in which those ballots were cast (``VBM'' are vote-by-mail ballots, 
``IP'' are ballots cast in the polling place and ``PRO'' are 
provisional ballots).
$b_p$ is the total reported ballots in that batch and Bound is the
upper bound on the discrepancy in the count for this group of ballots
(\textit{see} note~\ref{fn:1}). Yes and No are the total reported
votes for each selection and Audited indicates whether the set of
ballots was selected for the audit.
}
\label{tab:psov} 
\end{table}

The first post-election risk-limiting audit ever performed was
conducted by our group in Marin County in February of 2008 for Marin's
Kentfield School District Measure~A.
This ballot measure, passed by a {\small $2/3$} majority of voters,
raised property taxes in the Kentfield school district to support
public education.

Voters in 9~precincts were eligible to vote on Measure~A and 5,877
valid ballots were cast (280 showed undervotes and overvotes).
The initial vote count showed 4,216~votes (71.7\% of ballots) in favor 
and 1,661~votes (27.0\% of ballots)
against, with a margin of 298~votes (5.1\% of ballots) above the 
{\small $2/3$} majority of votes required for the measure to pass.
Table~\ref{tab:psov} summarizes the results for the Measure~A contest.

\subsubsection{Test \& Sample Size}
\label{sec:marin-feb08:test-statistic}

For this audit, error was measured as the overstatement of the margin,
in votes.
The method of~\cite{stark_conservative_2008} allows one to use a
weight function to accommodate factors such as an expected level of
discrepancy or variations in batch size.
We used the following weight function:\footnote{The notation
  $(\dots)_+$ means zero or the quantity in parentheses, whichever is
  larger.}
\begin{equation}
  \label{eq:marinwght}
          w_p(x) = \frac{(x-4)_+}{b_p},
\end{equation}
where $x$ is the overstatement of the margin in votes in batch $p$ and
$b_p$ is the total number of valid ballots cast in batch $p$.
This weight function ignores overstatements of up to 4~votes per
batch.
The risk calculation takes that allowance into account.

We set aside the smallest batch in a stratum of its own.\footnote{The
  excluded batch, precinct 2010, was a VBM-only batch in a precinct of
  6~registered voters.  We treated it as if it attained its maximum possible error,
  to ensure that the audit was conservative.}
  We used rolls of a 10-sided die to draw a
  simple random sample of 6 of the 8~batches of ballots cast
  in polling places to audit shortly after election day.
Once the vote-by-mail (VBM) ballots had been tabulated, we used rolls
of a 10-sided die to draw an independent simple random sample of 6 of
the 8~batches of VBM ballots to audit.
We postponed deciding whether to sample provisional ballots until we
could determine whether they could possibly change the outcome, given
the results of the audits of the polling-place and VBM ballots.
We thus had four strata containing a total of 18~batches: 
batches of ballots cast in polling places
(by precinct), batches of VBM ballots (by precinct) except for the smallest
precinct, the smallest VBM precinct by itself, and provisional
ballots.
By stratifying in this way we could start auditing polling-place
results almost immediately, even though VBM results were not available
until a couple of weeks after election day, and provisional results
not until the end of the canvass period.
Our protocol required a full hand count 
if we were unable to confirm the preliminary
results at our specified level of risk in the first round of sampling.
Table~\ref{tab:marin-b-schedule} shows the timetable for the audit.

\begin{table}[t]
\begin{center}
\begin{tabular}{l|l}
Milestone & Date \\
\hline
Election day & 5 February \\
Polling place results available & 7 February \\
Random selection of polling place precincts & 14 February \\
VBM results available & 20 February \\
Random selection of VBM precincts & 20 February\\
Hand tally complete & 20 February \\
Provisional ballot results available & 29 February \\
Computations complete & 3 March \\
\end{tabular}
\end{center}
\caption{ \protect \label{tab:timeline} Timeline for audit of Marin Measure~A, February 2008.
}
\label{tab:marin-b-schedule} 
\end{table}

\subsubsection{Risk Calculation}
\label{sec:marin-feb08:risk-calcs}

As shown in Table~\ref{tab:psov}, error in the provisional ballots
could have overstated the margin by up to 191~votes, and error in the
excluded precinct, precinct~2010, could have overstated the margin by
up to 4~votes.\footnote{\label{fn:1} To calculate the upper bounds in
  Table~\ref{tab:psov}, we assume that all invalid ballots and ``yes''
  votes might really have been ``no'' votes counted incorrectly,
  overstating the margin.  Counting a ``no'' vote as a ``yes'' vote
  overstates the true margin by 1~vote.  In contrast, counting a
  ``no'' as an invalid ballot or undervote overstates the margin by
  $2/3$ of a vote (since it subtracts a vote from both the numerator
  and the denominator of the margin calculation).  The upper bound on
  the amount by which error in the provisional ballots could have
  overstated the margin is thus the number of ``yes'' votes plus $2/3$
  of the number of invalid ballots: $176 + (2/3) \cdot (252-176-54) =
  190.67 \simeq 191$.  (For an extended discussion of how changing
  vote totals can affect election margins, \textit{see:}
  \url{http://josephhall.org/eamath/margins09.pdf}.)  }
At most, errors in these ballots could have inflated the apparent
margin over the true margin by 195~votes.
Unfortunately, any one of the other 16~batches---the 8~batches of
polling-place votes and 8~batches of vote-by-mail votes---could have
held enough error to account for the 103~vote ``reduced margin.''
Thus, only one batch among the 16 would have to have a margin
overstatement of more than 4~votes for the total overstatement in all
16 to possibly exceed 103~votes.

If only one of the batches had an overstatement of more than 4~votes,
then at least one of the polling-place counts or at least one of the
vote-by-mail counts had an overstatement more than 4~votes, or both.
If precisely one of the polling-place batches overstated the margin by
more than 4~votes, a random sample of 6 of 8~batches would have missed
it with probability\footnote{The notation ${x \choose y}$ is shorthand
  for the binomial coefficient $\frac{x!}{y!(x-y)!}$.}
\begin{equation}
  \label{eq:marin:detect}
  \frac{{7 \choose 6}}{{8 \choose 6}} = 25\%.
\end{equation}
By the same reasoning, if there were only one VBM batch with an
overstatement error of more than 4~votes, a sample of 6 of 8~batches
would have probability 25\% of missing it.
Since the chance of finding a single bad batch is at least 75\%
regardless of which stratum it was in, there is at least a 75\% chance
overall that the sample would contain the bad batch if there were only
one bad batch in all.
In other words, if exactly one of the 16~batches from which
the sample was drawn overstated the margin
by more than 4~votes, the chance the stratified sample of
12~batches would have missed it is 25\%.

Having only one bad batch is a hypothetical worst-case.
If two or more batches overstated the margin by more than 4~votes, the
chance that the sample would have missed all of them is considerably
less than 25\%.
The worst-case calculation guarantees that the risk is no greater than
25\%: For all other hypothetical situations in which the outcome is
wrong, the risk is lower.
None of the manual tally results had a discrepancy of more than
4~votes.
Hence, the audit limited the risk to at most 25\%, without a full hand
count.

The margin in this race is relatively small, about 4.8\% of the
ballots cast, including undervoted and invalid ballots.
The percentage of undervotes was about 4.5\%, larger than the margin.
And the race was small: only 9 precincts, of which one had only 6
registered voters.
These features made it necessary to audit a much higher percentage of
ballots than would have been necessary had the margin been larger, had
the race been larger, or had there been fewer undervotes.
The audit effort for this race is not typical: It is
virtually a worst-case scenario.
The other three pilot audits described below are of larger contests;
the required sampling fractions are correspondingly smaller.

The total cost of the manual tally was \$1,501, including the salaries
and benefits of four people tallying the count, a supervisor, and the
support staff needed to print reports, resolve discrepancies,
transport the ballots and locate and retrieve VBM ballots from the
batches in which they were counted.
This amounts to \$0.35 per ballot audited.
The tally took $1\frac{3}{4}$ days of counting to complete.

These figures do not include the statistician's time, most of which
was spent re-keying elections results from election management systems
(EMS) reports into machine-readable form.
That was not terribly burdensome because the contest was so small.
Performing the risk calculations took very little time.

\subsection{Yolo County, Measure W, November 2008}
\label{sec:yolo-08}

The second audit we performed was of Measure~W in Yolo County,
California.
Measure~W raised property taxes for the Davis Joint Unified School
District.
The 36,418 ballots cast contained 33,415 valid votes.
In all, 25,297~Yes votes (69.5\% of ballots) were cast, 
8,118~No votes (22.3\% of ballots) were cast, 
and 3,003~ballots (8.2\% of ballots) showed undervotes or
overvotes.
The measure passed with a margin of 17,179~votes
(51.4\% of valid ballots).

The race included 57~precincts.
For each precinct, VBM ballots were tabulated and reported separately
from IP ballots, giving 114~auditable batches.
The distribution of total votes per batch was centered at around
300--400 votes, with a handful of precincts with 100 or fewer votes.
Table~\ref{tab:yolo} reports details of the batches selected for
audit.

\subsubsection{Test \& Sample Size}
\label{sec:yolo:test-statistic}

\begin{table}[t]
\begin{center}  
\begin{tabular}{lrr|rr}
Batch ID & $b_p$ & Bound & Yes & No \\
\hline
100037-IP  & 396 & 594 & 285 & 87 \\
100039-VBM & 435 & 690 & 337 & 82 \\
100051-IP  & 443 & 600 & 280 & 123\\
100056-IP  & 284 & 437 & 209 & 56 \\
100060-IP  & 671 & 1001 & 483 & 153\\
100063-VBM & 356 & 548 & 257 & 65 \\
\end{tabular}
\end{center}
\caption{\protect 
Summary of audit sample for Yolo's Measure~W, November 2008.  
Six batches were selected using a random selection from the
57~decks of vote-by-mail ballots and
57~batches of ballots cast in polling places.  
``VBM'' denotes vote-by-mail ballots and ``IP'' denotes ballots
cast in polling places.
$b_p$ is the total reported ballots in that batch and Bound is the
upper bound on the discrepancy in the count for this group of ballots.
Yes and No are the votes for and against the measure.  
The audit found one error that increased the margin and one that
decreased it.
}
\label{tab:yolo} 
\end{table}

Like the audit of Marin County Measure~A described in the previous
section, this audit used a stratified random sample.  However, it used
the \textit{maximum relative overstatement} of pairwise margins (MRO)
of~\cite{stark_sharpdiscmeas_042008} instead of the margin
overstatement.\footnote{In contrast, the audits in Santa Cruz and
  Marin counties described below used unstratified sampling with
  probability proportional to size.}

The MRO divides the overstatement of the margin between each apparent
winner and each apparent loser by the reported margin between them.
In contests with more than two contestants, the MRO leads to a sharper
test than the raw margin overstatement used in the Marin Measure~A
audit, because it normalizes errors onto a scale of zero to 100\% of
the margin: An overstatement of one vote of a large margin (e.g.,
between the winner and fourth place) casts less doubt on the outcome
of a contest than an overstatement of one vote of a small margin
(e.g., between the winner and the runner-up).
In the Yolo election this does not matter because there were only two
candidates, ``yes'' and ``no.''

As in the audit of Marin Measure~A, batches consisted of votes for one
precinct cast in one way---in the polling place or by mail---but
provisional ballots were counted along with the polling-place ballots
for each precinct.
Moreover, we stratified the batches differently: Batches with
very small error bounds were grouped into one stratum.
The remaining batches comprised a second stratum.
The first stratum was not sampled.
Instead, batches in that stratum were treated as if they were attained
their maximum possible error.
A simple random sample was drawn from the second stratum.
We did not draw the sample until preliminary vote counts had been
reported for all batches in the county and provisional ballots had
been resolved.

As described in~\cite{stark_cast_022009}, batches can be exempted from
sampling if they are treated as if they attained their maximum
possible error.
This can improve efficiency if the worst-case error in those batches
is much smaller than in other batches.
That can happen if the batches contain relatively few ballots, as is
typical in rural precincts.
When such batches are set aside, a sample of a given size from the
remaining batches has a higher chance of containing a precinct that
holds a large error, if there is enough error in the aggregate to
alter the apparent outcome of the race.

In the Yolo county audit, we grouped 11~small batches that could
contain no more than 5~overstatement errors each into one stratum and
treated them as if they attained their worst-case error: a total of
0.04\% of the margin.\footnote{For the Yolo audit, we supposed that
  errors of up to 5~votes per batch might occur even when the outcome
  is correct; the Yolo County election staff had not seen an error of
  more than 1 or 2~votes in quite some time.  That led to grouping
  batches with error bounds of 5~votes or less into the unsampled
  stratum.  The resulting expected sample size was 2,121~ballots.  The
  actual sample size turned out to be 2,585~ballots.}
The remaining 103~batches comprised the second stratum.
By sampling from the second stratum alone, a sample of a given size
had a larger chance of finding at least one batch with a large
overstatement error---if there was enough error in all to cause the
apparent outcome to differ from the outcome a full hand count would
show.
There is a trade-off: Grouping more small batches into the unsampled
stratum entails assuming that there is more error, since those batches
must be treated as if they have their maximum possible error, to keep
the analysis conservative.
That means that less error can be tolerated in the remaining batches,
which reduces the error threshold that leads to expanding the audit.
However, it increases the sampling fraction among the remaining
batches, increasing the chance that the sample will contain a batch
with large errors---if any such batch exists---which tends to reduce the
initial sample size.

\subsubsection{Risk Calculation}
\label{sec:yolo-risk-calc}

To limit risk to 25\% required auditing an initial sample of 6 of the
103~batches in the second stratum.
The Yolo County Registrar of Voters Office randomly chose 6~batches
from those~103.
Auditing those batches, which contained 3,347 ballots, revealed two errors: 
One was a single vote
overstatement, and one was a single vote understatement.
This was below the pre-specified level that would trigger an expansion
of the sample, so the audit stopped without a full count.

Of the 6~batches selected for audit, one had already been hand counted
as part of the California 1\%~PEMT.
A team of three volunteers\footnote{Two of them were co-authors
  Miratrix and Stark.}  and one election official hand counted the
remaining 5~batches in approximately 4~hours.
Importing and editing data from the elections management system (EMS)
and performing the statistical calculations took considerably longer
than the hand counting.\footnote{This does not include the time
  required to write and debug the (re-usable) software.}

\subsection{Marin County, Measure~B, November 2008}
\label{sec:marin-08}

\begin{table}[t]
\begin{center}
\begin{tabular}{lrr|rr}
Batch ID & $b_p$ & $u_p$ & Yes   & No\\
\hline 
031-VBM & 91    & 0.009 & 49    & 30\\
043-VBM & 108   & 0.011 & 58    & 36\\
104-VBM & 40    & 0.004 & 22    & 13\\
191-VBM & 217   & 0.022 & 117   & 72\\
255-VBM & 246   & 0.025 & 133   & 81\\
286-VBM & 258   & 0.026 & 139   & 85\\
301-VBM & 245   & 0.025 & 132   & 81\\
339-VBM & 248   & 0.025 & 134   & 82\\
1002-IP & 316   & 0.018 & 151   & 110\\
1017-IP & 362   & 0.021 & 186   & 133\\
3013-IP & 277   & 0.015 & 125   & 102\\
3014-IP & 498   & 0.030 & 256   & 152\\
3017-IP & 318   & 0.018 & 154   & 111\\
3020-IP & 123   & 0.007 & 64    & 39\\
\end{tabular}
\end{center}
\caption{\protect 
Marin Measure~B audit results. 
Fourteen batches were selected at random with replacement
from 355~decks of vote-by-mail ballots (the 8~batch IDs ending in ``VBM'') 
and 189~batches of ballots cast in polling places 
(the 6 batch IDs ending in ``IP'')
using probability proportional to $u_p$.  
$b_p$ is the total number of ballots in the batch.  
$u_p$ is an upper bound on the maximum relative overstatement 
of the margin in the batch.  
Yes and No are votes for and against the measure.
The audit found no errors.
}
\label{tab:marin}
\end{table}

The third contest we audited in 2008 was Marin County's Measure~B,
which established two appointed positions, a director of finance and a
public administrator.
Measure~B was county-wide, so voters in all 189~precincts in Marin
County were eligible to vote.
A total of 121,295~ballots were cast: 61,839 Yes votes (50.1\% of ballots), 
42,047 No votes (34.7\% of ballots), and 17,409~undervotes and overvotes.
The margin was 19,792~votes (19.1\% of valid ballots).
Table~\ref{tab:marin} gives information about the batches selected for
audit.

\subsubsection{Test \& Sample Size}
\label{sec:marin-nov08:test-statistic}

This audit used the trinomial
bound~\cite{miratrix_trinomial_032009}.\footnote{The same method was
  used in the Santa Cruz audit described below.}
The risk limit for this audit was 25\%.

The trinomial bound is based on the \textit{taint} of the batches.
\textit{Taint} is the ratio of the MRO in a batch to the maximum
possible MRO of the batch.
Let $e_p$ denote the MRO of batch $p$ and let $u_p$ denote the maximum
possible MRO in batch $p$.
Then the taint of batch $p$ is $t_p \equiv e_p/u_p \le 1$.

To use the trinomial bound, taints of the batches in the sample are
compared to a pre-specified threshold $d$.
Each batch is in one of three categories: non-positive taint, taint up to
$d$, or taint greater than or equal to $d$.
The trinomial bound is based on the number of sample batches in each
category.

The trinomial bound uses weighted sampling with replacement rather
than SRS or a stratified random sample.
In each draw, the probability of drawing batch $p$ is proportional to
$u_p$.
This is called \textit{probability proportional to error bound} (PPEB)
sampling, since $u_p$ is a bound on the error in batch $p$.
Since the PPEB sample is drawn with replacement, batches can be
selected more than once.
For PPEB sampling, the joint distribution of the number of taints in
the categories is trinomial~\cite{miratrix_trinomial_032009}, and a
test of the hypothesis that the apparent outcome differs from the
outcome a complete hand count would show can be constructed using the
trinomial distribution.
The trinomial bound is closely related to the multinomial bound used
in financial auditing.
It is efficient when many auditable batches have no error or margin
understatements, some have small taints, and very few have large
taints.\footnote{The Kaplan-Markov Martingale approach described
  in~\cite{stark_pvalues_022009} seems to be at least as
  efficient and easier to compute.  We had not discovered the
  Kaplan-Markov Martingale approach when we conducted these audits in
  November~2008.}

Two practical considerations constrained the design of this
audit.
First, Marin County tallies VBM ballots in ``decks'' that are not
associated with geography.
To audit these ballots by precinct would have required either sorting
them or picking through a large number of decks to find the ballots
that corresponded to the precincts in the sample.
Either approach would have been prohibitively expensive and prone to
human error.
However, since the contest was countywide, every ballot included the
contest.
Hence, we could use decks of ballots as batches, without sorting them
by precinct.

Second, although the election management system (EMS) Marin County
uses can report the number of ballots in each deck, it cannot report
vote subtotals by deck.\footnote{In order to audit a batch of ballots,
  the auditors need an EMS report for the batch that lists
  contest-specific subtotals for all the ballots in the batch.}
That complicated calculating the bound $u_p$ on the maximum
relative overstatement of the margin in batch $p$.
We assumed the worst: that every vote in each deck was reported for
the apparent winner, but was in fact cast for the apparent loser.
If the EMS had been able to report subtotals by deck, we could have
used much smaller error bounds, and the initial audit sample size
would have been rather smaller.

\subsubsection{Risk Calculation}
\label{sec:marinB-risk-calc}

There are a variety of ways one might choose the initial number of
draws $n$ and the threshold $d$.
Based on the typical size of errors the election officials found in
past audits,
we set $d = 0.038$ and $n=14$;
see~\cite{miratrix_trinomial_032009} for more detail.

The choice of $d$ and $n$ does not affect the risk limit---$d$ and $n$
were chosen so that if the errors turned out to be like those seen in
previous audits, the audit could stop without enlarging the sample.
But whether errors are like those seen previously or not, the test has
at least a 75\% chance of requiring a full hand count if
the outcome is wrong, no matter what $n$ and $d$ are: It is guaranteed
to limit the risk to 25\% or less.

Since the PPEB sample is drawn with replacement, the same batch can be
drawn repeatedly, and the number of distinct batches can be smaller than
the number of draws.
For the Marin Measure~B audit, the expected number of distinct batches
was
\begin{equation}
  \label{eq:marin-expbatches}
   \sum_p \left( 1- \left( 1-\frac{u_p}{U} \right)^n \right) = 13.8.
\end{equation}
The expected number of ballots in those batches was
\begin{equation}
  \label{eq:marin-expballots}
  \sum_p b_p \left (1 - \left (1 - \frac{u_p }{U} \right )^n \right )
  = 3,424.
\end{equation}

When the error bounds $\{u_p\}$ vary considerably, the number of
batches one must audit using PPEB auditing methods tends to be smaller
than for SRS methods, when the outcome is correct.
Even though PPEB tends to select larger batches, the savings in the
number of ballots audited can still be dramatic.
For example, if we had used the method described for Yolo County to
audit this contest, the initial sample size would have been 22~batches
and the expected number of ballots to audit would have been 4,941,
about 44\% more than with PPEB and the trinomial bound.

To select the sample, the Marin County Registrar of Voters used
10-sided dice to produce a 6-digit random number.
We fed that 6-digit ``seed'' into the Mersenne Twister pseudorandom
number generator in the R statistics package, which we used to select
$n=14$ pseudo-random batches with replacement, with selection
probabilities proportional to the error bounds $u_p$.
The audit found no errors.
Hence, the audit could stop without expanding the sample, and the
outcome could be certified with a risk no greater than 25\%.

The total cost of this manual tally was approximately \$1,723 or
\$0.51 per ballot, requiring one team of four staff and one supervisor
2~days of work to pull the chosen ballots and count them by hand under supervision.
Performing the statistical calculations was not very time-consuming,
but pre-processing the EMS preliminary election results into a form
that could be used for statistical computations took several hours.

\subsection{Santa Cruz County, County Supervisor 1st District, November 2008}
\label{sec:sc-08}

\begin{table}[t]
\begin{center}
\begin{tabular}{lrr|rr|rr|rrr}
                           &           &              & \multicolumn{2}{|c|}{ Leopold } &  \multicolumn{2}{c|}{ Danner} 
       & & &  \\
Batch ID                & $b_p$ & $u_p$ & Reported      & Actual        & Reported & Actual     & MOV & $t_p$   & Times \\
\hline 
1002-VBM        & 573   & 0.28  & 251   & 252   & 227   & 227   & -1    & -0.002        & 1 \\
1005-IP         & 556   & 0.32  & 292   & 304   & 166   & 170   & -8    & -0.012        & 1 \\
1005-VBM        & 436   & 0.23  & 208   & 208   & 150   & 150   & 0     & 0             & 1\\
1007-IP         & 692   & 0.40  & 367   & 382   & 205   & 216   & -4    & -0.005        & 1\\
1007-VBM        & 630   & 0.33  & 311   & 311   & 240   & 240   & 0     & 0             & 1\\
1013-VBM        & 557   & 0.28  & 261   & 261   & 216   & 216   & 0     & 0             & 2\\
1017-VBM        & 399   & 0.21  & 191   & 191   & 139   & 139   & 0     & 0             & 1\\
1019-IP         & 448   & 0.25  & 218   & 223   & 128   & 137   & 4     & 0.007         & 1\\
1019-VBM        & 378   & 0.20  & 186   & 186   & 128   & 128   & 0     & 0             & 1\\
1027-VBM        & 232   & 0.11  & 107   & 107   & 98    & 98    & 0     & 0             & 1\\
1028-VBM        & 365   & 0.15  & 136   & 137   & 174   & 174   & -1    & -0.003        & 1\\
1037-VBM        & 758   & 0.33  & 261   & 261   & 309   & 309   & 0     & 0             & 2\\
1053-VBM        & 18    & 0.01  & 10    & 10    & 4     & 4     & 0     & 0             & 1\\
1060-IP         & 322   & 0.17  & 142   & 145   & 105   & 108   & 0     & 0             & 2\\
1073-VBM        & 20    & 0.01  & 11    & 11    & 3     & 4     & 1     & 0.036         & 1\\
1101-IP         & 721   & 0.35  & 312   & 321   & 275   & 279   & -5    & -0.007        & 1\\
\end{tabular}
\end{center}
\caption{\protect
Audit Data for Santa Cruz County Supervisor, 1st~District. 
The major contestants were Leopold and Danner; there were 103~votes for write-ins.  
Sixteen batches were sampled using PPEB, three of them twice.  
The number of ballots initially reported for batch $p$ is $b_p$.  
The upper bound on the MRO in batch $p$ is $u_p$.  
In each PPEB draw, the probability of selecting batch $p$ is 
proportional to $u_p$.  
MOV is number of votes by which error changed the apparent margin for Danner.  
The taint $t_p$ is the observed overstatement of the margin in the batch divided
by the maximum possible overstatement of the margin in the batch.
Times is the number of times the batch was selected in 19~PPEB draws.
Two positive taints were found, both less than the threshold
$d = 0.047$ that would require expanding the audit.
}
\label{tab:santaCruz}
\end{table}

The fourth contest we audited was Santa Cruz County Supervisor,
1st~District.
This contest spanned 76~precincts in which 26,655~ballots were cast,
including (at the time of the audit) 
12,103~votes (45.4\% of ballots) for John Leopold and 
9,964~votes (37.4\% of ballots) for Betty Danner, the runner-up.
Leopold had the plurality, winning by a margin of 2,139~votes
(8.0\% of ballots cast; 9.6\% of valid votes).
Table~\ref{tab:santaCruz} lists reported and audited votes for each
candidate for the batches in the sample, along with other statistics
of the audit.

\subsubsection{Test \& Sample Size}
\label{sec:sc-nov08:test-statistic}

The Santa Cruz audit used the trinomial bound, as
discussed above for Marin Measure~B.
Table~\ref{tab:santaCruz} lists the precincts audited, the errors
found, and the corresponding taints.
There were $n=19$~PPEB draws, resulting in 16~distinct batches containing
7,105~ballots.
Three of the batches were selected twice; their audit results enter
the calculations twice, even though they only need to be hand counted
once.

The trinomial bound allowed a much smaller sample size than SRS would
have.
If we had used a simple random sample, the initial sample size would
have been 38 batches containing, in expectation,
13,017~ballots---almost double the PPEB sample size.
The savings is larger than in the Marin Measure~B audit because some
batches could hold errors of up to $49\%$ of the margin.
As a result, enough error to change the outcome could have been hidden
in as few as two batches.
With SRS, every batch has the same chance of being selected and so the
chance of auditing such a batch is low.
With PPEB sampling, batches that can hold particularly large errors
have particularly large chances of being audited.
Hence, a smaller sample suffices---unless it revealed
large taints.

\subsubsection{Risk Calculation}
\label{sec:sc-nov08:risk-calc}
We set $n = 19$ and $d = 0.047$; \cite{miratrix_trinomial_032009}~describes
how we chose those values.

While analyzing the data from the manual tally we learned that the
hand tally had included provisional ballots, while the batch totals on
which we had based the audit calculations did
not.\footnote{Apparently, 806~provisional ballots had been cast in the
  race in all.  Among the audited batches, precinct~1005 had 37; 1007
  had 30; 1019 had 32; 1060 had 11; and 1101 had 39.}
Accordingly, the number of ballots in several batches in the sample
increased and margins in those batches changed.
The audit also found differences of one ballot in some VBM batch
totals.
We attribute those differences to ballots that needed special
treatment to be tabulated properly.\footnote{For example, cases where
a voter used ``X-marks'' instead of filling in the ovals next to their
choice.  If the X-mark is not centered, the optical scanner might not
consider the voting target dark enough to be a valid vote.  During the
hand tally, voter errors where intent is clear can change the
unofficial results on which the audit calculations are based.}
To ensure that our calculations were conservative, we treated every
change to the reported margins---including changes produced by
provisional ballots and ballots that might have required special
treatment---as error in the reported hand count, i.e., as error
revealed by the audit.\footnote{It would also have been conservative
  to treat all the provisional ballots as error, but we had no way to
  separate the votes for the provisional and original ballots in the
  audit, so it was impossible to isolate the error in the original
  counts.}
 
The largest observed taint, 0.036, was an overstatement of one vote in
a very small batch.
The largest overstatement, 4~votes, was in a much larger batch; the
resulting taint was only 0.007.
When the test is based on taint, errors of a vote or two in small
batches can have a large influence, because they can translate to
large taints.
However, small batches are relatively unlikely to be sampled using
PPEB, because they have small values of $u_p$.

Error that changed the margin in either direction was
as large as 8~votes in some batches.
Based on past experience, this is an unusually high error rate.
As far as we can tell, this discrepancy was simply miscommunication
about the provisional ballots, not error in the counts \textit{per
se}; nonetheless, we treated it as error, to be conservative.

Changing $b_p$, the number of ballots in batch $p$, affects $u_p$, the
upper bound on the MRO for the batch.
If $u_p$ is still larger than $e_p$ for every batch $p$, the audit
remains statistically conservative.
Since there were so few provisional ballots and the bound $u_p$ is
quite conservative in the first place---it is calculated by assuming
that all the votes in batch $p$ should have gone to the loser---it is
highly implausible that $e_p > b_p$ in any batch.
However, this experience emphasizes how important it is for 
auditors and election officials and their staff to communicate clearly.

Most of the provisional ballots in the sample turned out to be votes
for the winning candidate, Leopold, so they increased the margin,
strengthening the evidence that the apparent outcome was correct.
In fact, only two batches had positive taints, both less than the
threshold $d = 0.047$,\footnote{The category counts were 17~batches
  with non-positive taint, 2~with positive taint below $d$, and none
  with taint above $d$.} so the audit could stop after the initial
sample.
Risk was limited to 25\%, without a full hand count.

The manual count took approximately 3 days at a total cost of \$3,248,
or \$0.46 per audited ballot.\footnote{Three
  precincts were randomly drawn twice.  Because these precincts
  only need to be hand counted once, they are only counted once in the
  per-ballot cost figure cited here.}
The audit team consisted of one supervisor and four counting staff and
the work included pulling ballots, hand counting them, recording the
counts, and compiling the count data for the Official Statement of the
Vote.
Performing the statistical calculations did not require much time, but
translating preliminary election results from EMS output into a form
amenable for calculations took several hours.
In all four pilot audits, the inability of commercial EMS to export
data in useful formats added considerably to the difficulty of
election auditing.

\section{Discussion}
\label{sec:discuss}

We performed four rigorous risk-limiting audits on races of different
sizes with different margins using different sampling designs,
different ways of defining batches of ballots, different ways of
stratifying batches, and different statistical tests.
The cost and the time required were modest.
Basing audits on samples drawn with probability proportional to an
error bound (PPEB) can be far more efficient than simple random
sampling or than stratified random sampling using strata based on the
mode of voting (in the polling place versus by mail) when the number
of ballots per batch varies widely.
There remains room for big gains in efficiency---that is, for reducing
the number of ballots that must be counted to confirm an electoral
outcome that is, in fact, correct.

Risk-limiting audits are currently feasible for only a few races at a
time; however, Stark~\cite{stark_collections_052009}, extends the MRO
in a way that allows a collection of contests to be audited efficiently as
a group.
We have also developed more efficient sequential tests since these
pilots were performed.
We expect to test those methods in November~2009 and June~2010.
How to combine stratification and PPEB remains an open theoretical
question that will need to be addressed to use PPEB to audit contests
that cross jurisdictional boundaries.

We set the risk limit to 25\% in these four audits in order to control
the work involved in these experiments: Our primary goals were to test
the feasibility of the methods and to gain experience, not to limit the
risk to a very low level.
Nothing in the methods demands a high limit: We could have
chosen a smaller limit, at the cost of more hand counting.
But for very small contests, limiting the risk to, say, 1\%, will
generally require counting nearly every ballot by hand.

Risk of 25\% translates to a ``confidence level'' of
75\%.\footnote{This is not technically a confidence level as the term
  is used in Statistics, but it is consistent with how the term is
  used in election auditing.}
We suspect some election integrity advocates would not be satisfied if
this value were mandated by legislation.
We do not advocate any particular limit on risk: Choosing the risk
limit is a job for policymakers.
However, we feel that it is better to guarantee a modest risk limit
rigorously---knowing that the risk is almost surely much lower---than
to claim a lower risk limit on the basis of \textit{ad hoc},
untestable assumptions, such as WPM of 20\%.

Similarly, a method that deals with error rigorously in deciding
whether to expand the audit is far preferable to one that stops
whether the audit finds error or not.
We advocate using rigorous, conservative methods to determine the risk guarantee
of an audit method.
The risk of a conservative method can then be evaluated under more optimistic
assumptions, such as WPM of 20\%.
That allows statements of the form, ``the risk is guaranteed to be no greater than
10\%.  And if a WPM of 20\% holds, the risk is no greater than 1\%.''

Efficient risk-limiting audits are complicated and difficult for the
public to understand.
Designing them requires statistical expertise that we suspect is rare
among the staff of elections officials.
Given the problems brought to light by studies such as the
California TTBR~\cite{casos_ttbr_2007}, which showed that voting
systems have inadequate computer and physical security controls,
developing in-house statistical expertise is a lower priority than
developing better security and chain of custody.
However, we hope to provide turn-key procedures and open-source
software for risk-limiting audits, and these pilots helped us
understand how to make a procedure efficient,
comprehensible, and comprehensive.

A particularly time-consuming step in the pilot audits was translating
batch-level data into machine-readable formats.
The Election Management Systems (EMS) in the counties we worked with
are oriented towards displaying tables of results, not computing.
For instance, when election results are exported from EMSs in
comma-separated value (CSV) format, columns do not line up, values are
out of place, and headers are repeated.
A great deal of scripting and hand editing was required to make the
exported data useful.
In some cases, we double-keyed data by hand from reports.
The scripting and data editing introduces opportunities for error 
and makes it impractical to audit even a modest number of large 
races in a short canvass period.

Election auditing requires better ``data plumbing'' than EMS
vendors currently provide.
We hope EMS vendors will improve their products to export
structured data.
One suitable format is the OASIS Election Markup Language (EML), which
wraps election data in a descriptive container much like XHTML, a
structured version of HTML, the language that powers the World Wide
Web.\footnote{\textit{See:}
  \url{http://www.oasis-open.org/committees/election}.  EML is the
  work of OASIS' Voter \& Election Services Technical Committee (TC).
  Author Hall is a member of this TC. }
Using structured data facilitates accurate import and export of data
across systems and makes it easy to perform statistical computations
and generate any report users might want.

EMSs are also a bottleneck for auditing batches smaller than
precincts.
Generally, the fewer ballots in each auditable batch, the less
hand counting is required overall when the outcome is correct.
Moreover, logistical considerations can make it desirable to audit
ballots cast in polling places before ballots cast by mail have been
tabulated, as we did in the Marin Measure~A audit.
Hence, it would help if EMSs could report subtotals by batch,
keeping different ballot types---and even ballots cast on different
machines---separate.
Currently, most cannot.

As we described above, this EMS limitation complicated the audit of
Marin Measure~B and increased the workload.
To get batch totals for the VBM batches in the Marin Measure~B audit
required a labor-intensive kludge: On a non-production copy of the
database, every deck but one was deleted from the report.
That manual process was repeated for each VBM batch in the sample.
Needless to say, this was tedious and error prone.

Moreover, it was only practical to do this for the batches in the
sample, not for every batch.
That made it necessary to use extremely pessimistic error bounds on
the decks.
This in turn increased the sample size---and the workload.

The limitations of EMSs also affected the audit of Santa Cruz's County
Supervisor race.
There, the problem was that the EMS combines all ballot types for a
given precinct and reports results without distinguishing among them.
Provisional ballot totals are combined with in-precinct ballots
totals in the reports; there is no way to separate them for the
purpose of the audit.
That required us to treat changes to totals that had nothing to do
with miscount (and only with the fact that provisional ballots are
counted later in the canvass than other ballots) as error.

Sorting out the situation in Santa Cruz took some time, which
emphasizes the importance of clear communication between auditors and
election officials.
We understood that provisional ballots were not included in the
initial totals.
We intended that they not be included the audited totals for the
batches selected, so that we could make apples-to-apples comparisons.
However, we learned while analyzing the data that the hand count
totals included provisional ballots.
That required us to re-think how we were dealing with provisional
ballots and how we were defining error.
Fortunately, we were still able to confirm the outcome rigorously
without expanding the audit.

On the whole, we believe that it is premature to advocate
risk-limiting audits for many races simultaneously.
Auditing methods are developing quickly, but they need to be more
efficient to be practical on a wide scale.
Improving EMS ``data plumbing'' and developing step-by-step auditing
guides for elections officials are crucial as well.

\section{A Modest Proposal for Risk-Limiting Audits}
\label{sec:modprop}

Methods that do not have a built-in procedure for going to a full
count cannot limit risk.
The method we describe in this Section is not a method that we would
advocate, but it shows that it is possible to limit risk without
complicated calculations involving the errors the audit finds.
Indeed, this method doesn't require any computation at all.
It has a known chance of progressing to a full hand count when the
outcome is wrong, and the same known chance of progressing when the
outcome is right.

Risk-limiting audits that use statistical computations involving the
observed discrepancies to decide when to stop counting are complex.
Even with help from professional statisticians, the logistical hurdles
are high.
In light of these difficulties, we propose a radically simple
risk-limiting audit: Count a given race by hand in its entirety with
some pre-specified probability, regardless of the rate of errors found
by the audit.
This simple scheme is risk-limiting because it has a pre-specified
probability of a full count if the outcome is wrong (of course, it has
the same chance of proceeding to a full count if the outcome is right,
which makes the approach inefficient statistically).

This simple scheme does not take into account any features of the
contests themselves, but it is easy to embellish it.
For instance, we might want to have a higher chance of counting close
races by hand, and we might want to count some batches from every
race, to ensure that all races receive some scrutiny.
These goals can be met without any on-the-fly statistical
calculations.
For instance, an audit could have three components:
\begin{itemize}
\item \textit{Basic Audit Level:} A fixed percentage of batches from
  every race is hand counted.  If that count reveals many errors, the
  entire race is counted by hand.  This provides quality control, but
  has a chance of correcting wrong outcomes.  The basic level of
  auditing could be set very low, e.g., $0.5\%$ of batches, to limit
  workload.
\item \textit{Full Recount Trigger:} Any contest with a sufficiently
  small margin is counted by hand in its entirety.  The margin that
  triggers the full count could be geared to the accuracy of the
  original tabulation technology.
\item \textit{Random full hand count:} Every race has some positive
  probability of being counted by hand entirely.  That probability
  could depend on a variety of things, including the size of the race
  and the margin of the race.  One possible functional form is
  \begin{equation}
    \label{eq:modaud}
    P_r = \frac{f_r}{20} + \frac{1}{1000 \cdot m_r},
  \end{equation}
  where $P_r$ is the probability that race $r$ is fully counted
  manually, $f_r$ is fraction of registered voters eligible to vote in
  the race (the number of eligible voters in the race divided by the
  total number of eligible voters for the election) and $m_r$ is the
  margin in the race expressed as a fraction.
  
  For statewide races $f_r$ is $1$.  For the formula above, such a
  race would have at least a 5\% chance of a full hand count; the
  chance would increase as the margin shrinks.  A statewide race with
  a 1\% margin would have a 15\% chance.  A local race in which 0.5\%
  of voters can vote and that has a 5\% margin would have a 2\%
  chance of a full hand count.
\end{itemize}
The third component makes this proposal truly risk-limiting, because
it ensures that every race has a known, minimum chance of a full hand
count whenever the outcome is wrong.
It is not efficient, because races have the same chance of a full hand
count when the outcome is right.
But the alternative---sequential testing---requires statistical
calculations during the audit, which may remain impractical for most
jurisdictions.

This proposal is a straw man.
In particular, the functional form and the constants are arbitrary.
This proposal explicitly trades counting efficiency for simplicity
within a minimally risk-limiting framework.
But the general approach could be tuned to address practical concerns,
such as overall workload.

\section{Conclusion}
\label{sec:conc}

Current audit laws and proposed legislation do not control the risk
that an incorrect outcome will be certified.
We have tested four variations of risk-limiting audits on contests of
various sizes in California, using different ways of drawing samples,
different ways of defining batches of ballots, different ways of
stratifying batches, different ways of quantifying error, and
different statistical tests.
The cost of these audits was nominal, on the order of tens of cents
per audited ballot,\footnote{The average cost of both the Marin audits and
  the Santa Cruz audit reported here was \$0.44.} and they required
small teams a few days to complete.

Even though these risk-limiting audits were inexpensive, rigorous and
logistically manageable, the methods are complex: Simpler methods
might be preferable, even if they require more hand counting.
To that end, we proposed an extremely simple approach to risk-limiting
audits.
The approach limits the risk by ensuring that every race has a
strictly positive probability of being fully counted by hand.
It guarantees that every race gets some auditing for quality control
and that races with margins below the ``noise level'' of the counting
technology get counted by hand completely.
The downside is that the approach is statistically inefficient: It
requires more counting than necessary when the apparent outcome is
correct.

The consensus definition of ``risk-limiting'' audits requires
controlling the chance of error in each contest separately: Whenever
the apparent outcome is incorrect, there must be a high chance of
catching and correcting the error by requiring full manual count.
Limiting a different measure of risk, such as the long-run fraction of
certified outcomes that are certified erroneously, could decrease the
hand-counting burden of auditing substantially.
However, methods that control this ``false electoral rate'' are likely
to be at least as complex statistically as those we tested here, and
hence may not be practical for routine auditing.

Ballot-level auditing, as described
by~\cite{johnson_vvpataudit_102004, neff_confidence_122003,
calandrino_machineassist_082007}, may lead to more efficient
risk-limiting audits.
There are barriers to ballot-level auditing, including associating a
given physical ballot (or ballot record) with an electronic ballot
record in a one-to-one manner without compromising ballot secrecy.
Moreover, while ballot-level auditing can greatly reduce the amount of
hand counting required, the calculations to limit risk are nearly as
complex as for larger batch sizes.

Publishing ballot images, as the Humboldt County Elections
Transparency Project did, also holds promise for ensuring the accuracy
of elections.
But this too poses problems that have not yet been addressed.
For example, there needs to be a provision for auditing the
completeness and accuracy of ballot images.
And there needs to be a way to ensure that ballots cannot be
associated with individual voters, to prevent vote selling or
coercion.

Election auditing has developed remarkably over the past five years.
We are hopeful that within the next five years there will be methods
that are rigorous, simple and efficient enough to be used universally.

\section*{Acknowledgments}
\label{ack}

This work was possible only because of the generous and tireless
dedication of the election staff and volunteers of Marin, Santa Cruz
and Yolo Counties.

We are grateful to Sean Flaherty, Mark Halvorson, Mark Lindeman, Neal
McBurnett, John McCarthy, Eric Rescorla, Pam Smith, Howard Stanislevic
and anonymous referees for discussion and comments on earlier drafts.

Hall was supported in this work by the National Science Foundation
under A Center for Correct, Usable, Reliable, Auditable and
Transparent Elections (ACCURATE), Grant Number \mbox{CNS-0524745}.
Miratrix was supported in this work by a National Science
Foundation Graduate Research Fellowship (Fellowship~ID:
2007058607).

Any opinions, findings, and conclusions or recommendations expressed
in this material are those of the authors and do not necessarily
reflect the views of the National Science Foundation.

\bibliographystyle{amsplain}
\bibliography{draft.bib}

\end{document}